\documentclass[12pt]{article}
\input{epsf}
\title{\bf    Electroweak   radiative  effects
in  the  single  $W$-production \\
at Tevatron and LHC }
\begin{document}
\author{I. Akushevich$^{a)}$\footnote{on leave
of absence from the National
Center of Particle and High Energy Physics,
220040 Minsk, Belarus}, A.Ilyichev$^{b)}$, N.Shumeiko$^{b)}$,
V.Zykunov$^{c)}$
}
\date{}
\maketitle
\vspace*{-12mm}
\begin{center}
{\small {\it $^{(a)}$ North Carolina Central University,
Durham, NC 27707, USA \\ and  \\
Jefferson Lab, Newport News, VA 23606, USA\\}}
{\small {\it $^{(b)}$ National
Center of Particle and High Energy Physics,
220040 Minsk, Belarus \\}}
{\small {\it $^{(c)}$
GSTU, 246746 Gomel,  Belarus
\\}}
\end{center}

\begin{abstract}
An alternative calculation of the lowest order electroweak
radiative corrections (EWC) to the single $W$-boson
production in hadron-hadron  collision
in the framework of the quark parton model
without any absorption of the collinear quark singularity
into the parton distributions
is carried out.
Numerical analysis under Tevatron and LHC
kinematic conditions is performed.
\end{abstract}

\section{Introduction}

To define the experimental value of the $W$-boson mass $m_W$ with
a high precision
it is necessary to take into account the radiative effects \cite{snow2}.
Such calculations for the single $W$-production at hadron-hadron
collider experiments already have been done. So in Ref.~\cite{FSR}
radiative effects from the final state photon emissions
to the single $W$-production were investigated.
Basing on more accurate calculation of
EWC to the $W$-production in a general 4-fermion process
\cite{WH} the lowest order corrections in hadronic collisions
have been calculated in Refs. \cite{pp,pp2}, where the contributions of
both initial and final state radiations have been taken into account.
The total EWC to the polarization observables of $W$-boson production at RHIC
within the covariant approach \cite{covar} were estimated by one of us (V.Z.)
in Refs. \cite{epj,yaf}. The multiphoton radiation in leptonic $W$-boson decays
was found in a recent paper \cite{mph}.

 In the present report the new explicit formulae for EWC to the
inclusive single $W$-production in hadron-hadron collisions are presented.
For extraction of infrared singularity the covariant method
\cite{covar} was used. It is well-known that
the modern fits to the quark distributions
do not include the EWC.
That is why,
like Refs. \cite{epj,yaf},
performing the calculation
we leave the quark mass singularity without any changes,
therefore in spite of Refs. \cite{pp,pp2}
our results are essential depended on the quark masses.

This report is organized as follows. In Sect.~2 the Born
approximation for
the single $W$-production is presented. In Sect.~3 the total lowest
order EWC are considered. The detail numerical results are
available in Sect.~4. The conclusions are given in Sec.~5.

\section{Kinematics and Born approximation}

The process of the single $W$-boson production in hadron-hadron
collision
\begin{equation}
p+p (\bar p) \rightarrow W^{\pm }+X \rightarrow
l^{\pm }+X
\label{main}
\end{equation}
can be described by two pairs of the quark-antiquark
subprocesses according to the charge of $W$-boson.
For the $W^-$ production we have the processes
\renewcommand{\theequation}{2.\alph{equation}}
\setcounter{equation}{0}
\begin{equation}
 q_i(p_1)+{\bar q}_{i'}(p_2) \rightarrow  W^{- }\rightarrow
l^{-}(k_1)+{\bar \nu}_l(k_2),
\label{3w}
\end{equation}
\begin{equation}
 {\bar q}_i(p_1)+ q_{i'}(p_2) \rightarrow  W^{-}\rightarrow
l^{-}(k_1)+ {\bar \nu}_l (k_2);
\label{4w}
\end{equation}
and for $W^+$ ones read as
\setcounter{equation}{0}
\renewcommand{\theequation}{3.\alph{equation}}
\begin{equation}
 q_i(p_1)+{\bar q}_{i'}(p_2) \rightarrow  W^{+ }\rightarrow
l^{+}(k_1)+ \nu _l(k_2),
\label{1w}
\end{equation}
\begin{equation}
 {\bar q}_i(p_1)+ q_{i'}(p_2) \rightarrow  W^{+ }\rightarrow
l^{+}(k_1)+ \nu _l(k_2).
\label{2w}
\end{equation}
\renewcommand{\theequation}{\arabic{equation}}
\setcounter{equation}{3}
Here $p_1$, $p_2$, $k_1$, $k_2$ are the four-momenta of
corresponding particles ($p_1^2=m_1^2$, $p_2^2=m_2^2$,
$k_1^2=m_l^2$ and $k_2^2=0$).

The standard set of Mandelstam variables on
the quark-parton level reads
\begin{equation}
s=(p_1+p_2)^2,\; t=(p_1-k_1)^2,\; u=(k_1-p_2)^2,
\end{equation}
and for hadrons
\begin{equation}
S=(P_1+P_2)^2,\; T=(P_1-k_1)^2,\; U=(k_1-P_2)^2,
\label{mvh}
\end{equation}
where $P_1$ and $P_2$ are four-momenta of initial hadrons
($P_1^2=P_2^2=m_N^2$).

According to the quark parton model (QPM) the
substitution $p_i \rightarrow x_iP_i$ is used.
Here $x_i$ is the fraction of the first ($i=1$) or second ($i=2$) hadronic
momentum carried by the corresponding struck quark.
This procedure will be denoted by operator "hat".
As a result in QPM the invariants $s, t, u$ can be expressed via $S, T, U$
and $x_i$  as
$$ \hat s \approx x_1x_2S,\ \hat t \approx x_1T,\ \hat u \approx x_2U.$$

At the hadron-hadron collisions the center of parton-parton masses frame
has an undetermined motion along the beam direction.
Therefore the hadronic Mandelstam variables can be expressed through
the standard set of variables in following way:
$$ T=-\sqrt{S}|{k_1}_{\perp}|e^{-\eta},\;
   U=-\sqrt{S}|{k_1}_{\perp}|e^{\eta}, $$
where $\sqrt{S}$ is a centre-of-mass energy,
$|{k_1}_{\perp}| \equiv {k_1}_T$ is
a transversal to the beam
direction component of the detected particle four-vector
and $\eta$ is a pseudorapidity.

After integration over azimuth angle
($d^3k_1/{k_1}_0 \Rightarrow \pi d\eta d{k_1}_{\perp}^2$),
the hadron-hadron cross section can be
presented as the sum over all flavors of quarks and antiquarks
\begin{eqnarray}
\label{born}
 \frac{ d \sigma_B }{d\eta d{k_1}_{\perp}^2}
&=&\sum_{i,i'} \int dx_1 dx_2 f_i(x_1,Q^2) \Sigma
\delta(x_2+\frac {x_1T}{x_1S+U})
 = \sum_{i,i'} \int \limits^1_{x_1^m} dx_1 f_i(x_1,Q^2) \Sigma_0,
\end{eqnarray}
where
\begin{eqnarray}
\Sigma
  = \frac{ \pi \alpha^2}{4N_cs_w^4}
\frac { |V_{ii'}|^2\hat B_{ii'}
{\hat \Pi}_l
{\hat \Pi}_l^+
f_{i'}(x_2,Q^2)}
{{\hat s}(x_1S+U)},\;\;
\Sigma_0=\Sigma \arrowvert _{x_2=x_2^0},
\label{sig0}
\end{eqnarray}
and
\begin{equation}
\Pi_l=\frac 1 {s-m_W^2+im_W\Gamma _W}
\end{equation}
is $W$-propagator ($\Gamma_W$ is the $W$-boson width).
The lower limit of integration in (\ref{born}) reads $x_1^m=-U/(S+T)$,
and the other variables are defined in the following way:
$x_2^0=-x_1T/(x_1S+U)$, $N_c=3$ is the color factor,
$V_{ii'}$ is CKM matrix element,
$f_i(x_i,Q^2)$ are the spin averaged quark densities,
$Q^2$ is a typical transfer momentum square in the partonic reaction,
$B_{ii'}=u^2$ ($t^2$)
for (\ref{3w}) and (\ref{2w}) ((\ref{1w}) and (\ref{4w}))
subprocesses.

\section{The lowest order radiative corrections}
\begin{figure}[t]
\vspace{-2cm}
\begin{tabular}{cccc}
\begin{picture}(80,100)
\put(-15,-70){
\epsfxsize=6cm
\epsfysize=6cm
\epsfbox{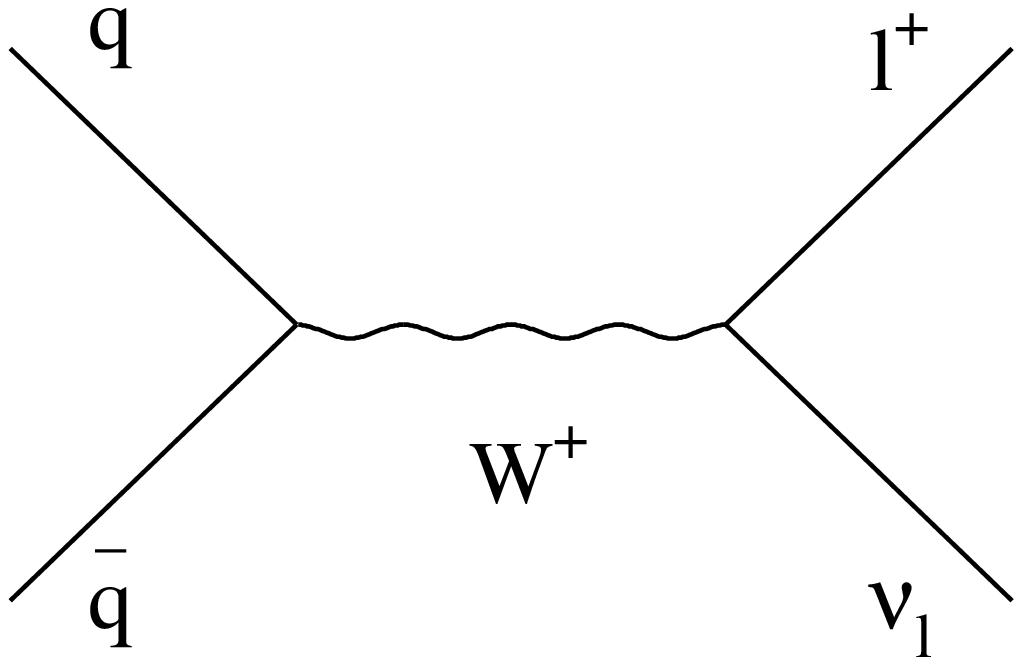} }
\end{picture}
&
\begin{picture}(80,100)
\put(0,-70){
\epsfxsize=6cm
\epsfysize=6cm
\epsfbox{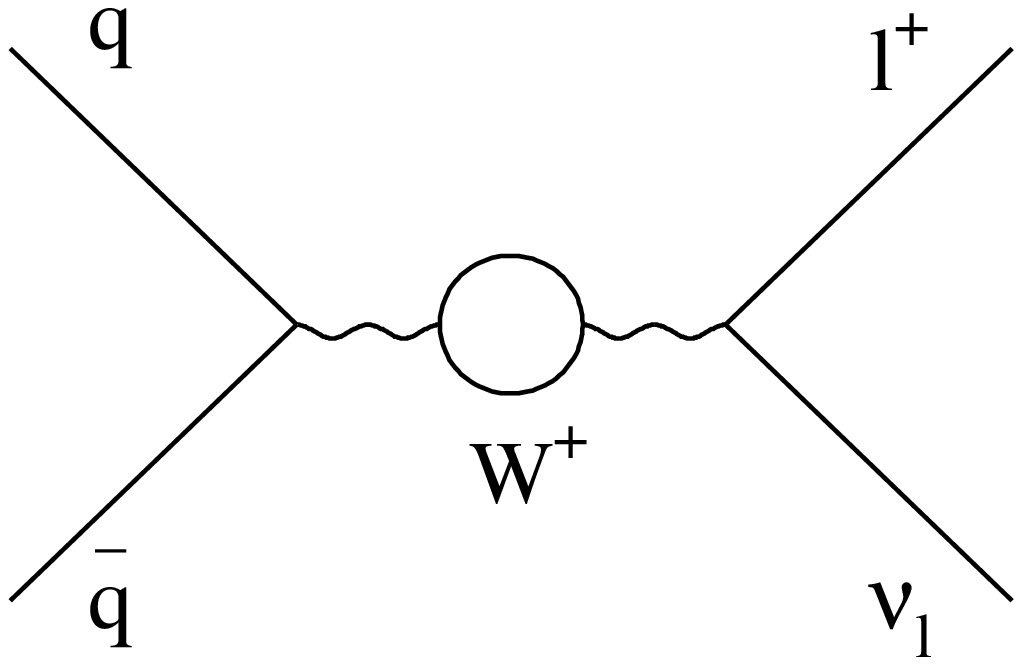} }
\end{picture}
&
\begin{picture}(80,100)
\put(15,-70){
\epsfxsize=6cm
\epsfysize=6cm
\epsfbox{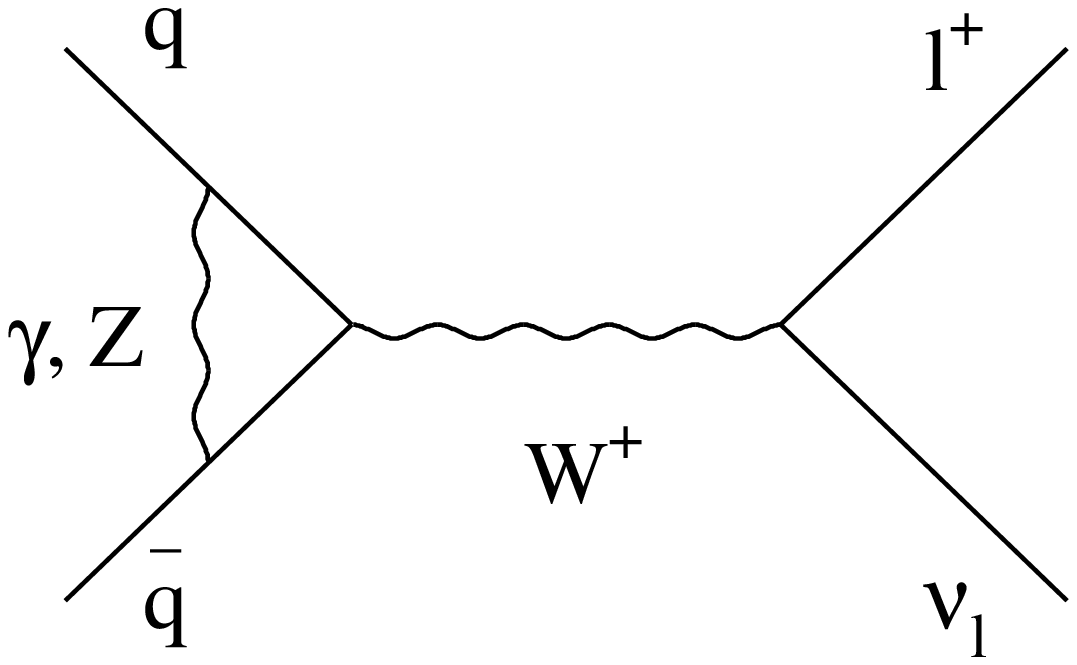} }
\end{picture}
&
\begin{picture}(80,100)
\put(30,-70){
\epsfxsize=6cm
\epsfysize=6cm
\epsfbox{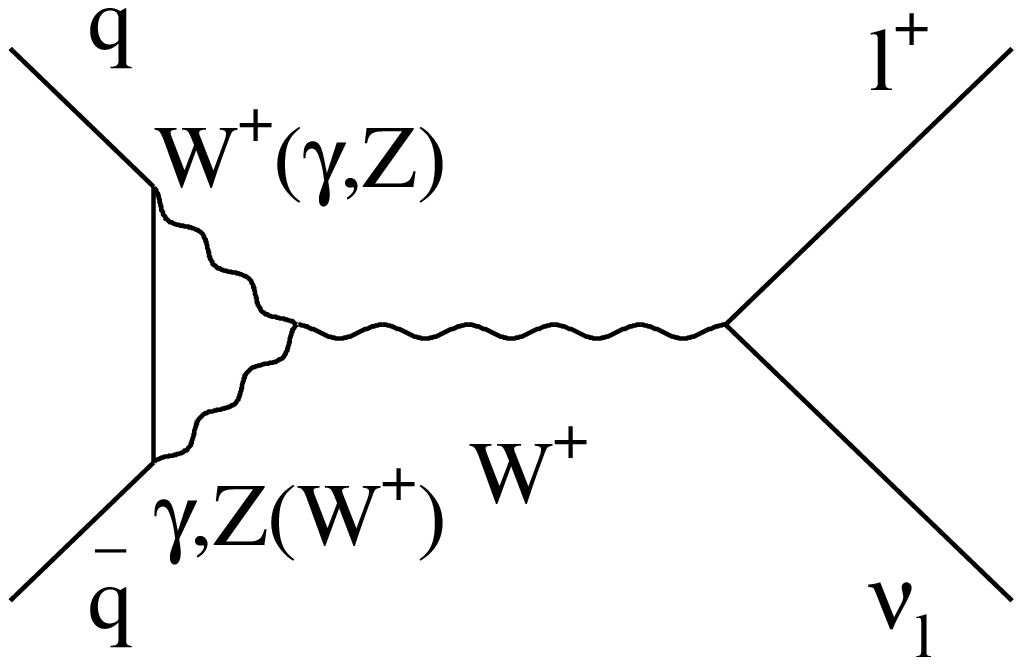} }
\end{picture}
\\[15mm]
\begin{picture}(80,100)
\put(-15,0){
\epsfxsize=6cm
\epsfysize=6cm
\epsfbox{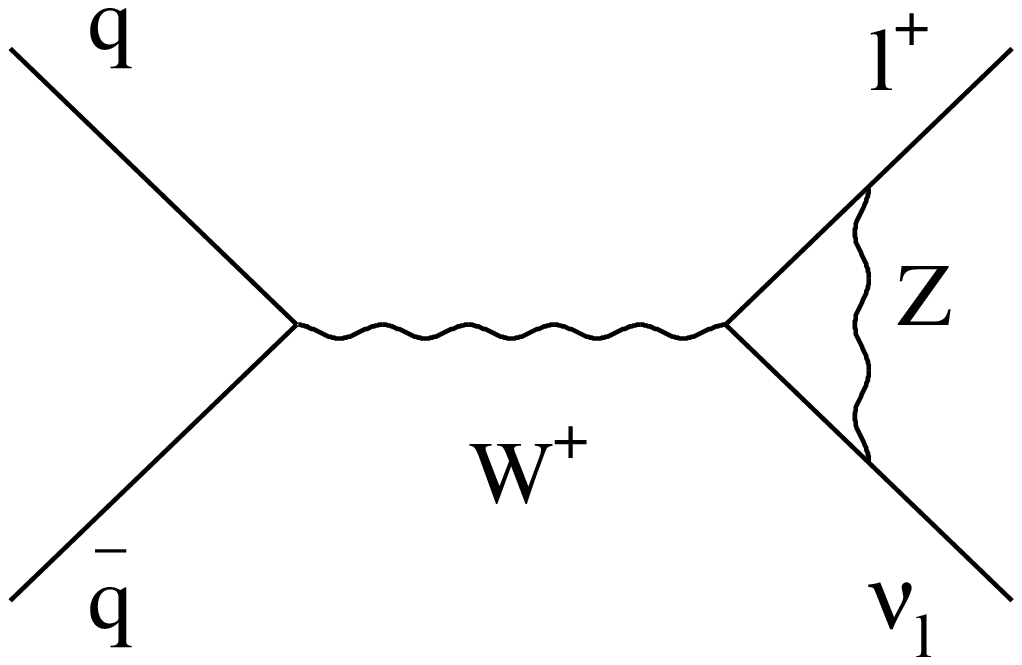} }
\end{picture}
&
\begin{picture}(80,100)
\put(0,0){
\epsfxsize=6cm
\epsfysize=6cm
\epsfbox{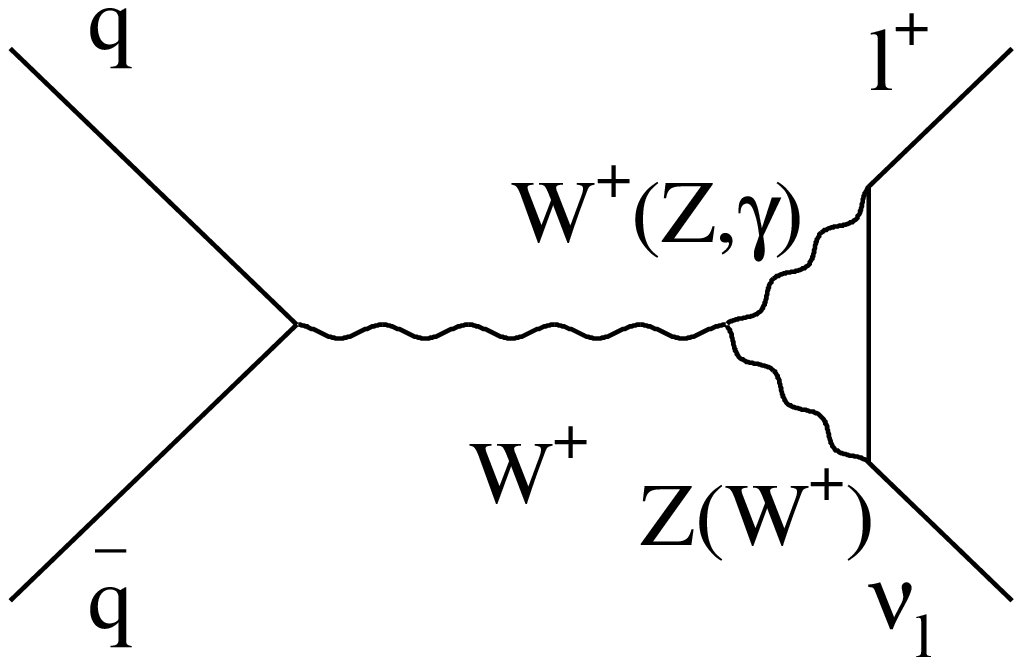} }
\end{picture}
&
\begin{picture}(80,60)
\put(15,0){
\epsfxsize=6cm
\epsfysize=6cm
\epsfbox{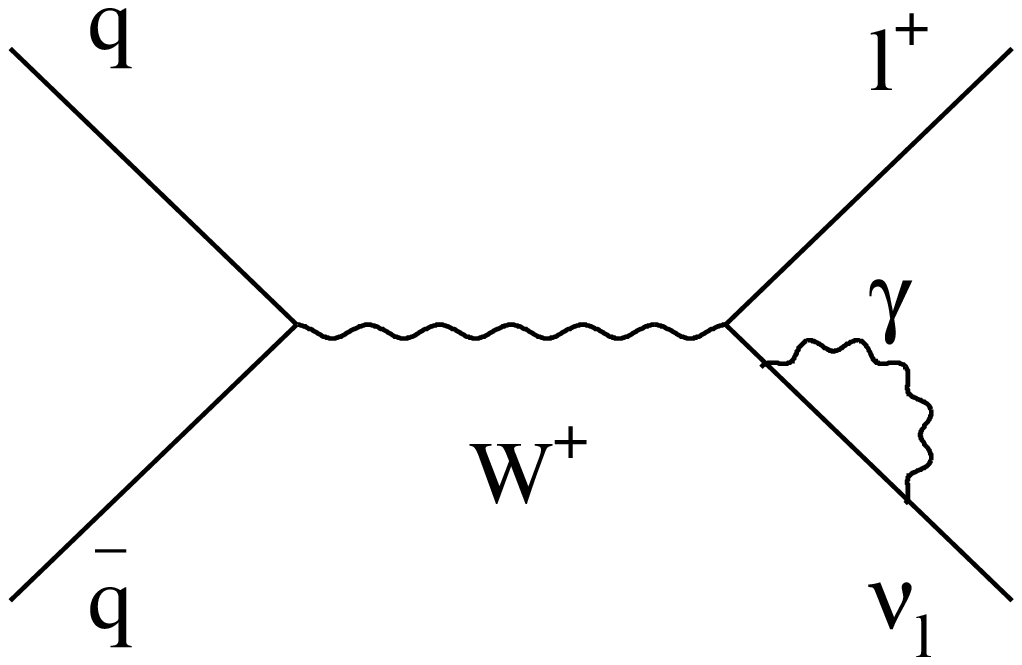} }
\end{picture}
&
\begin{picture}(80,100)
\put(30,0){
\epsfxsize=6cm
\epsfysize=6cm
\epsfbox{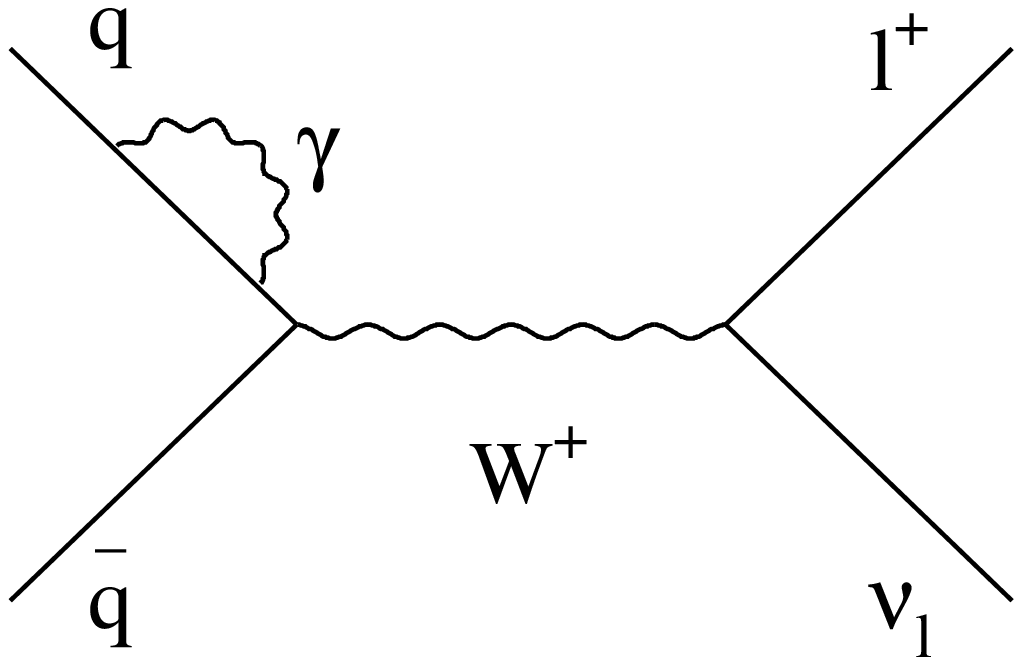} }
\end{picture}
\\[-10mm]
\begin{picture}(80,100)
\put(-15,0){
\epsfxsize=6cm
\epsfysize=6cm
\epsfbox{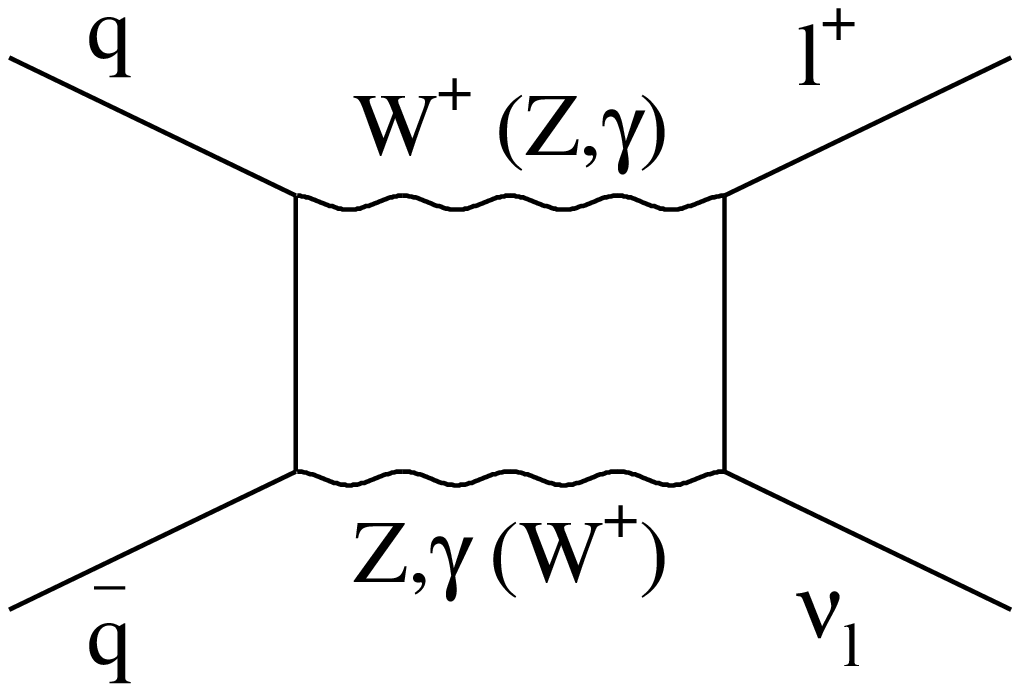} }
\end{picture}
&
\begin{picture}(80,100)
\put(0,0){
\epsfxsize=6cm
\epsfysize=6cm
\epsfbox{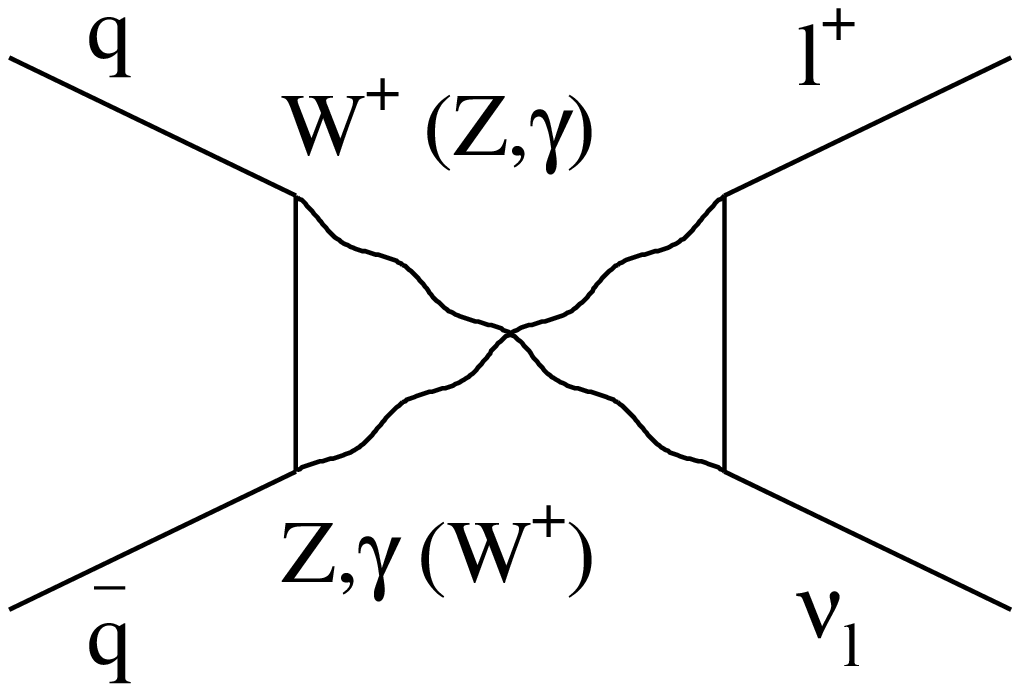} }
\end{picture}
\end{tabular}
\vspace{-15mm}
\caption{\protect\it
Born and
the virtual one-loop diagrams for $q \bar q \rightarrow l^+ \nu$
process. The contribution to the self-energy of $W$-boson
is symbolized by the empty circle.}
\label{1}
\end{figure}

\begin{figure}
\vspace{-1cm}
\begin{tabular}{cccc}
\begin{picture}(80,100)
\put(-15,-70){
\epsfxsize=6cm
\epsfysize=6cm
\epsfbox{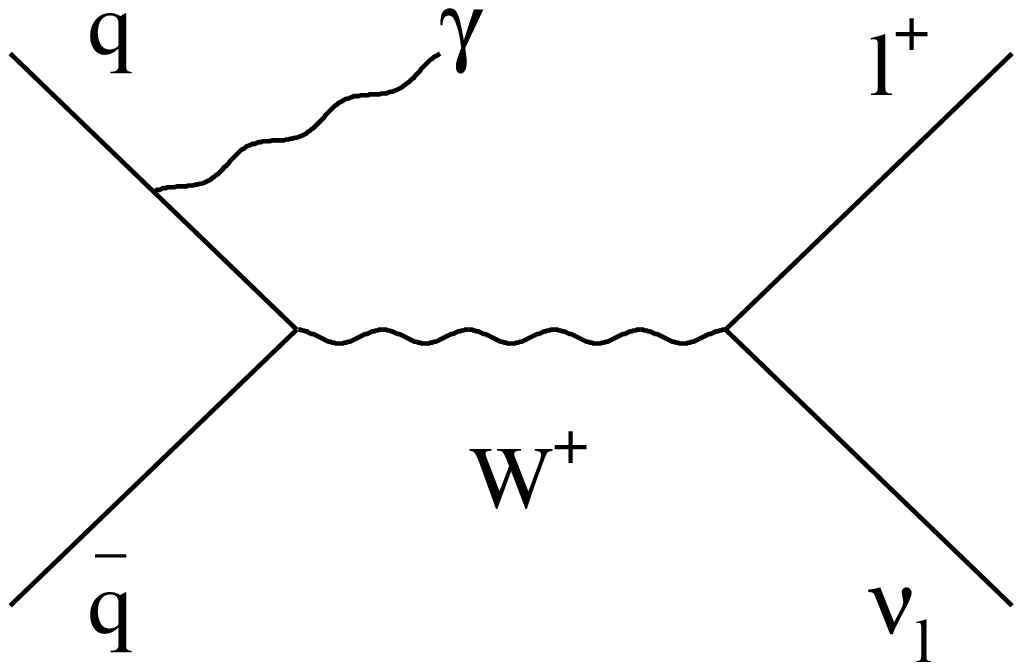} }
\end{picture}
&
\begin{picture}(80,100)
\put(0,-70){
\epsfxsize=6cm
\epsfysize=6cm
\epsfbox{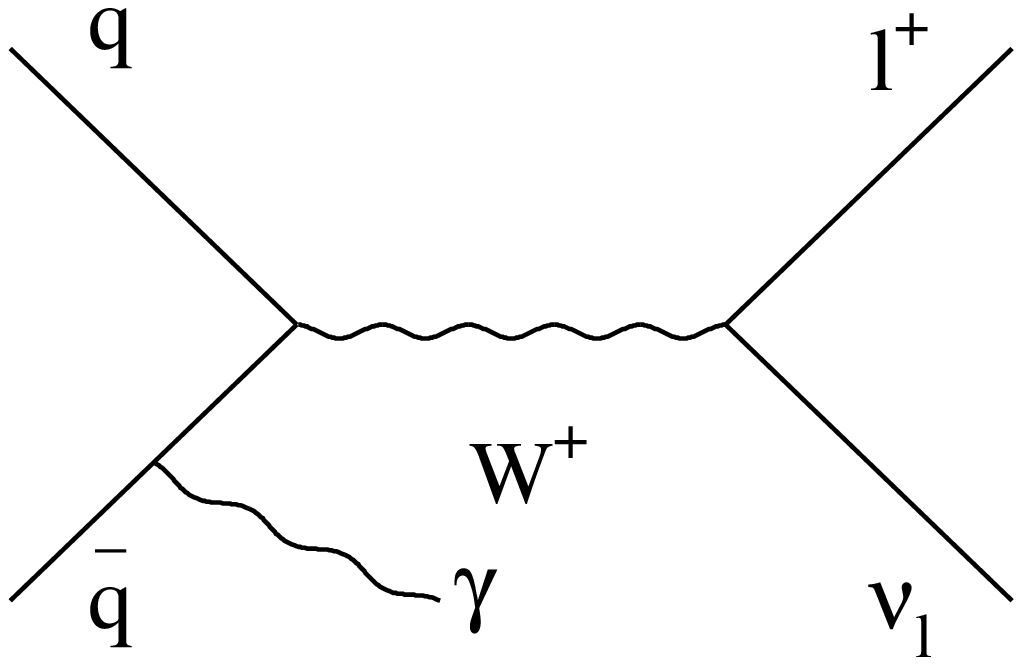} }
\end{picture}
&
\begin{picture}(80,100)
\put(15,-70){
\epsfxsize=6cm
\epsfysize=6cm
\epsfbox{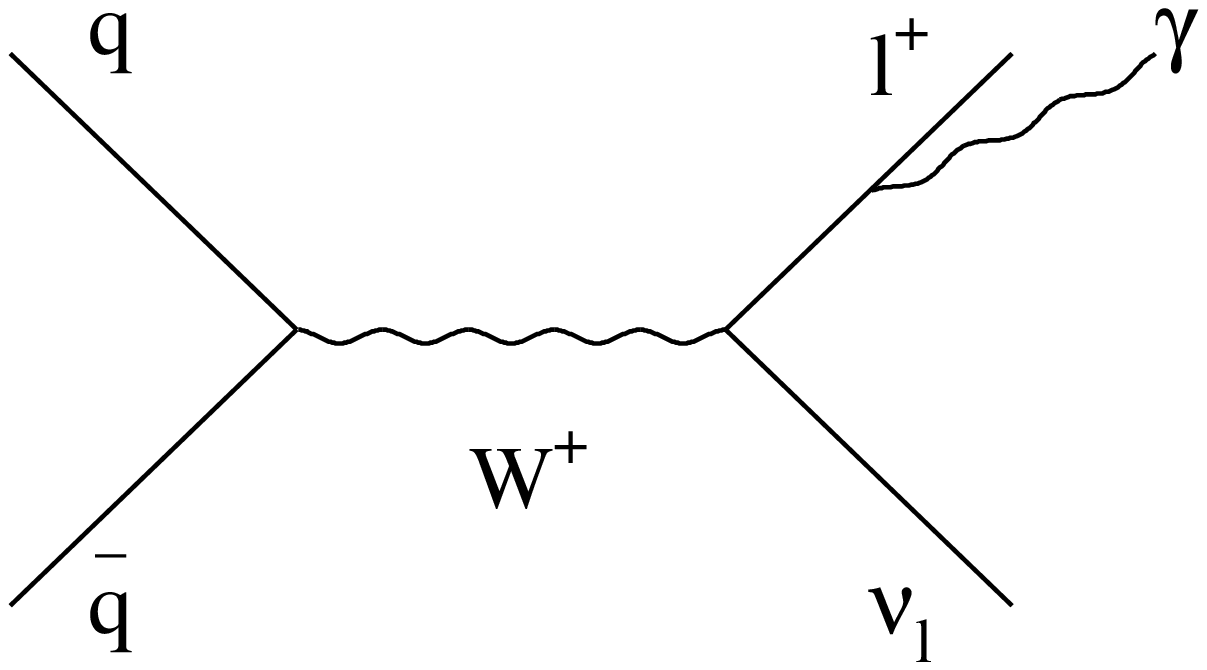} }
\end{picture}
&
\begin{picture}(80,100)
\put(30,-70){
\epsfxsize=6cm
\epsfysize=6cm
\epsfbox{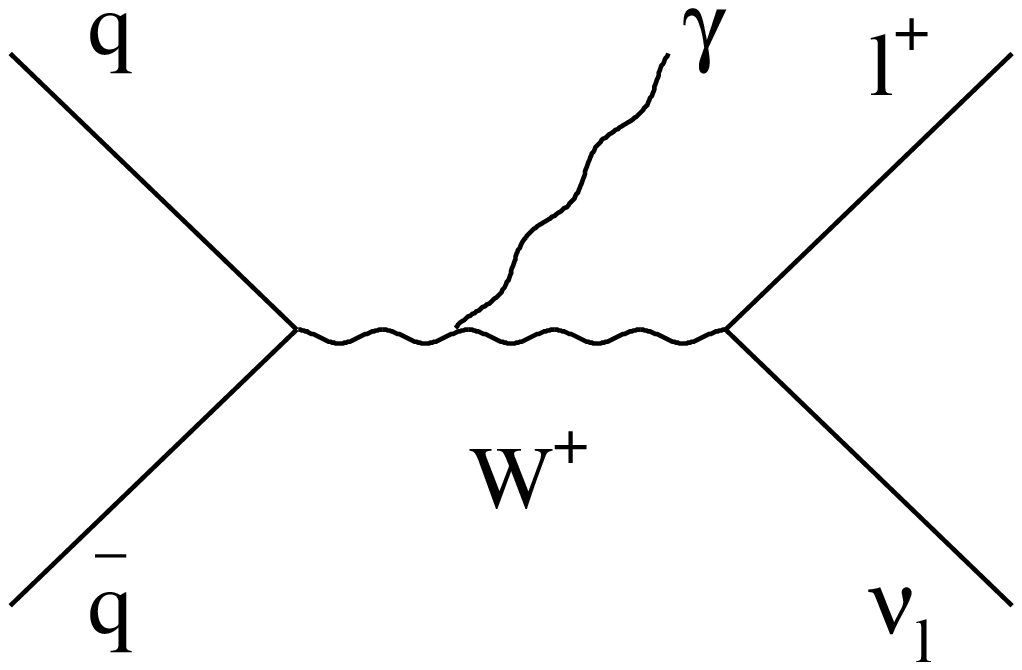} }
\end{picture}
\end{tabular}
\vspace*{10mm}
\caption{\protect\it
Bremsstrahlung diagrams for $q \bar q \rightarrow l^+ \nu \gamma$
process.}
\label  {2}
\end{figure}

The lowest order radiative corrections to the process (\ref{main})
can be presented as a sum of contributions appearing both from the additional
virtual particles (the V-contributions, see Fig. \ref{1})
and the real photon emissions (the R-contributions, see Fig.
\ref{2}):
\begin{equation}
\frac{ d \sigma_{RC} }{d\eta d{k_1}_{\perp}^2}=
\frac{ d \sigma_V }{d\eta d{k_1}_{\perp}^2}+
\frac{ d \sigma_R }{d\eta d{k_1}_{\perp}^2}.
\label{RC}
\end{equation}

Notice that the V- and R- contributions include infrared (IR)
divergences while the sum of them (\ref{RC}) has to be IR free.
Therefore one of the main aim of the radiative corrections
calculation consists in the cancellation of the IR divergences
correctly. Since as it will be presented below
the final expressions for
the V-contribution factorize in front of the Born cross section
and we integrate inside of this factor analytically,
the extraction of the IR-part from this contribution can be
performed
in a simple way. At the same time dealing with radiated process
we need to integrate over an unobservable particle (a photon) phase space.
Therefore for the extraction of the IR-part from the
R-contribution
it is necessary to make some manipulation with integrand
expression that will be presented below too.
Now consider the V- and R-contribution in details.

 The V-contribution
is proportional to the Born one
and could be written as
\begin{equation}
 \frac{ d \sigma_V }{d\eta d{k_1}_{\perp}^2}
 = \sum_{i,i'} \int dx_1 f_i(x_1,Q^2)
   \hat \delta^{ii'} _{V}|_{x_2=x^0_2}
   \Sigma_0.
\label{v}
\end{equation}
Here the factor $\delta^{ii'} _{V} $ consists of the seven terms
\begin{equation}
\delta^{ii'} _V =  \delta _W+\delta _{Vl} +\delta _{Vq} +
\delta _{Sl}+ \delta _{Sq} +\delta^{ii'} _{\gamma W}+\delta^{ii'} _{ZW},
\label{delta}
\end{equation}
which are contributions
of
the one-loop diagrams (see Fig.1)
to the cross section (\ref{main}):
the $W$-boson self-energy $\delta _W$;
the leptonic vertex $\delta _{Vl}$;
the quark vertex $\delta _{Vq}$;
the neutrino self energy $\delta _{Sl}$;
the up-quarks self energy $\delta _{Sq}$;
the $\gamma W$- and $ZW$  boxes
$\delta^{ii'} _{\gamma W}$
$\delta^{ii'} _{ZW}$ respectively. The explicit expressions for
each term can be found in Refs. \cite{BS1,BS2}.

The IR-part of the V-contribution can be presented
by the formula (\ref{v}), with the following
replacement $\delta^{ii'}_V \rightarrow \delta_V^{IR}$,
where
\begin{eqnarray}
\delta  ^{IR} _V&=&\frac {\alpha}{2\pi}\log \frac s {\lambda ^2}
(Q^2_l+Q^2_i+Q^2_{i'}
-Q_iQ_{i'}\log \frac {s^2}{m^2_i m^2_{i'}}
+Q_lQ_ic_l\log \frac {t^2}{m^2_im^2_l}
\nonumber \\&&
-Q_lQ_{i'}c_l\log \frac {u^2}{m^2_{i'}m^2_l} ),
\end{eqnarray}
$Q_j$ is the charge of the fermion $j$ expressed in the units of
the elementary charge $e=\sqrt{4\pi \alpha}$
(e.g. $Q_u=+2/3$, $Q_d=-1/3$), and  $c_l=+1$ (-1)
for (\ref{3w}) and (\ref{2w})  ((\ref{1w}) and (\ref{4w}))
subprocesses.
The rest part of the V-contribution to the cross section contains
finite correction
\begin{equation}
\delta^{ii'}_V - \delta^{IR}_V
 =\delta^{ii'}_V(\lambda^2 \rightarrow s).
\end{equation}

The contribution of bremsstrahlung
\begin{equation}
p+p (\bar p) \rightarrow W^{\pm }+X \rightarrow
l^{\pm }+\gamma+X
\label{brem}
\end{equation}
to the process (\ref{main})
after extraction of IR divergence by the method described in Ref.
\cite{covar} can be presented as a sum of the infrared dependent
and infrared free parts
\begin{equation}
\frac{ d \sigma_R}{d\eta d{k_1}_{\perp}^2}=
    \frac{ d \sigma_R^{IR} }{d\eta d{k_1}_{\perp}^2}
  + \frac{ d \sigma_R^F    }{d\eta d{k_1}_{\perp}^2}.
\label{sir}
\end{equation}

Summing up IR-parts of the V- and R- contributions we get
\begin{eqnarray}
\frac{d \sigma_R^{IR}}{d\eta d{k_1}_{\perp}^2}
+\frac{d \sigma_V^{IR}}{d\eta d{k_1}_{\perp}^2} =
   \sum_{i,i'} \int dx_1 f_i(x_1,Q^2)
   \frac{\alpha}{2\pi}\Sigma_0 {\hat J}(t,0)
\log \frac{ {\tilde v}^2 {\hat s}_0 }{v_{max}^2},
\label{cancel}
\end{eqnarray}
i.e. IR divergence has canceled successfully. Here
\begin{eqnarray}
\label{j}
&& J(t,v)=
 Q_l^2 - c_lQ_lQ_i \log \frac{t^2}{m^2_lm_1^2} +
 c_lQ_lQ_{i'} \log \frac{u^2}{m_l^2m_2^2}
 + Q_i^2 - Q_iQ_{i'}\log\frac{s^2}{m_1^2m_2^2}  + Q_{i'}^2.
\nonumber \\
&&v_{max}=x_1(S+T)+U,\;\;
\tilde v =  - \frac{Tx_1}{\sqrt{S}}(1+\frac{TU}{(x_1S+U)^2}).
\end{eqnarray}

After IR-terms extraction
the finite part of  the R-contribution (so called "hard" photon
contribution) reads
\begin{eqnarray}
 \frac{d\sigma^F_R}{d\eta d{k^2_1}_{\perp}}&=&
   \sum_{i,\ i'} \int dx_1dx_2 f_i(x_1,Q^2) f_{i'}(x_2,Q^2)
   {\hat \Sigma}^F_R.
\end{eqnarray}
The kinematic regions for integration over $x_1$ and $x_2$ are restricted by
the following region
   $$ -\frac{U+m_N^2-m_l^2}{S+T+m_N^2} \leq x_1 \leq 1, \ \ \
                   x_2^0 \leq x_2 \leq 1,$$
and $\Sigma ^F_R$ reads
\begin{eqnarray}
\Sigma^F_R
  =\frac{\alpha^3}{8N_cs_w^4s}|V_{ii'}|^2
         (Q_l^2\Pi_l\Pi_l^+V_l+Q_l\Re[\Pi_l]V_{lq}+V_q
        +Q_l\Pi _l\Pi
_l^+V_{lw}+\Re[\Pi_l]V_{qw}+\Pi_l\Pi_l^+V_w).
\end{eqnarray}
The indexes of the $V$-terms correspond to radiated leg
($l$ --- final lepton, $q$ --- initial quarks, $w$ --- $W$-boson),
as well as
the double index corresponds to the same interference term.
\begin{figure}[t]
\vspace*{37mm}
\begin{tabular}{cc}
\begin{picture}(120,120)
\put(-20,0){
\epsfxsize=9cm
\epsfysize=9cm
\epsfbox{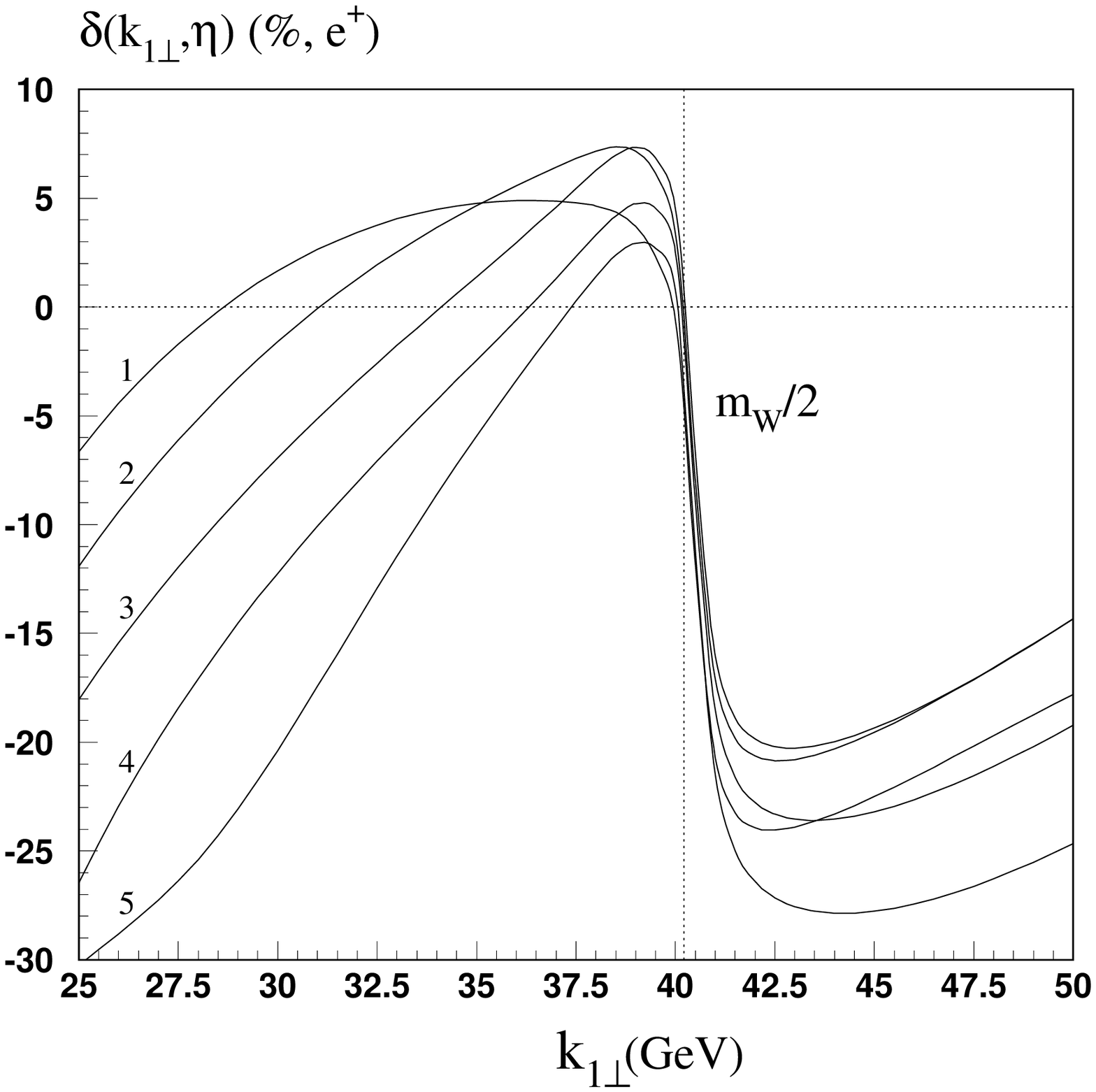} }
\end{picture}
&
\begin{picture}(120,120)
\put(90,0){
\epsfxsize=9cm
\epsfysize=9cm
\epsfbox{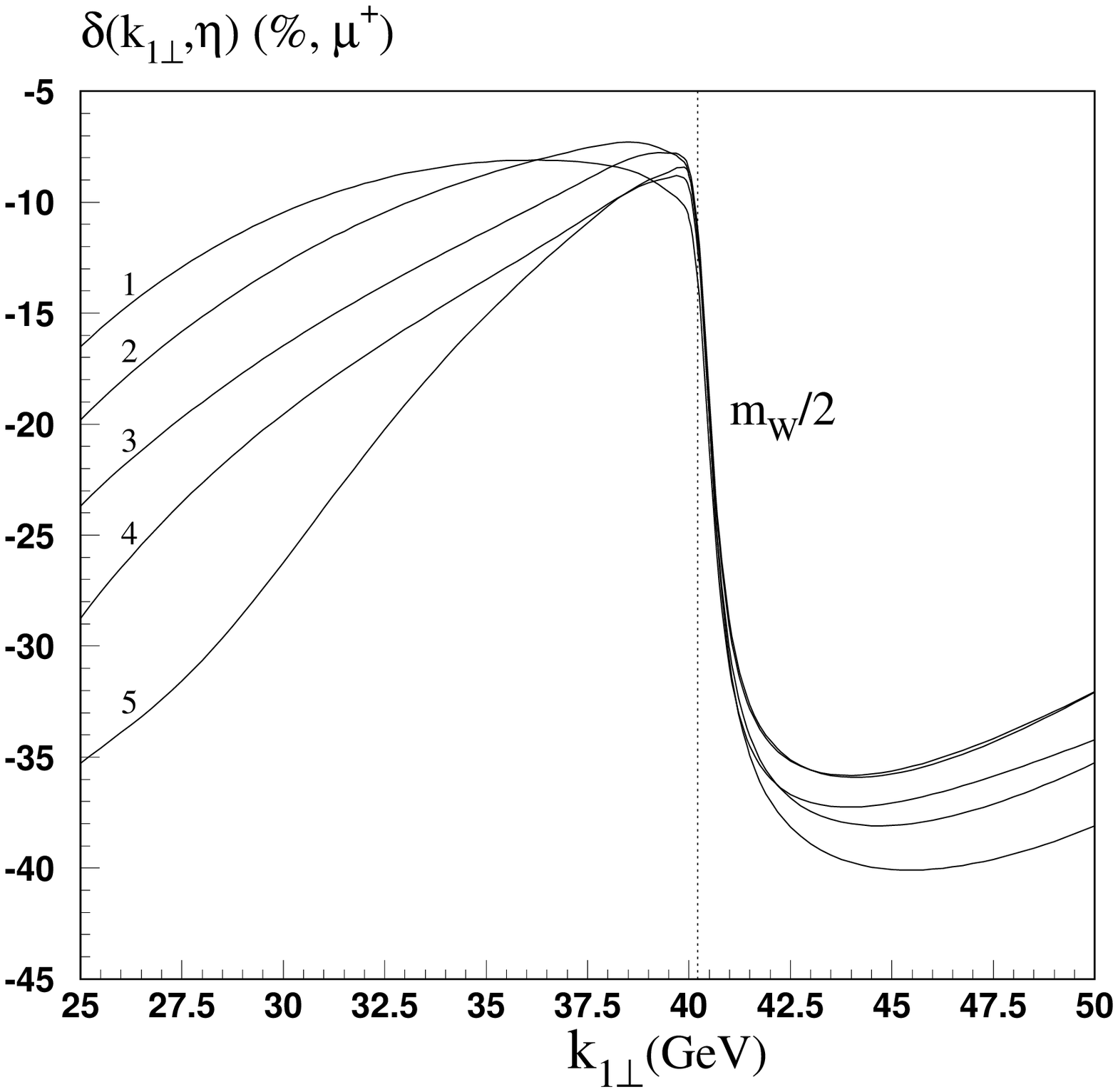} }
\end{picture}
\end{tabular}
\vspace*{-5mm}
\caption{\protect\it
Relative corrections to the lepton transverse momentum distribution
at $\sqrt{S}=1.8$~TeV, (Tevatron) for different pseudorapidity $\eta $.
The numbers $1,2,3,4,5$ correspond to the lines with $\eta=-2,-1,0,1,2$
respectively.
The quark masses are defined by (\ref{mm}).
}
\label{tkty}
\end{figure}
\section{ Numerical results}
The presented above formulae allows us to estimate numerically
the lowest order radiative effects for the $W$-production at the
Tevatron ($p\bar p$ collisions, $\sqrt{S}$=1.8TeV)
and LHC ($pp$ collisions, $\sqrt{S}$=14TeV)
kinematic conditions.

The performed analysis has shown that the differences between
the relative corrections which are defined as
a ratio of EWC to the Born contribution
\begin{equation}
\delta(k_{1\bot },\eta )=\frac
{d\sigma _{RC}/d\eta dk_{1\bot}}
{d\sigma _{B}/d\eta dk_{1\bot}}
\end{equation}
calculated within
MRS98 \cite{MRS}, GRV98 \cite{GRV}, CTEQ \cite{CTEQ} parton distributions
are not essential and MRS98 will be used for
the numerical analysis in this section
(we use as well as in Ref.\cite{pp} $Q^2=m_W^2$).

For the numerical analysis we use the rule of the naive QPM
$p_q=x_qP_N$, when the transversal component of the quark momentum
inside of the hadron can be dropped that leads to the following
relation
\begin{equation}
m_q=x_qm_{N},
\label{mm}
\end{equation}
where $x_q$ is an argument of corresponding
parton distribution. It should be noted that
numerical estimations of EWC with
this choice for quark masses and with
$m_u = 5$ MeV, $m_d = 9$ MeV
give
very close results.

\begin{figure}[t]
\vspace*{37mm}
\begin{tabular}{cc}
\begin{picture}(120,120)
\put(-20,0){
\epsfxsize=9cm
\epsfysize=9cm
\epsfbox{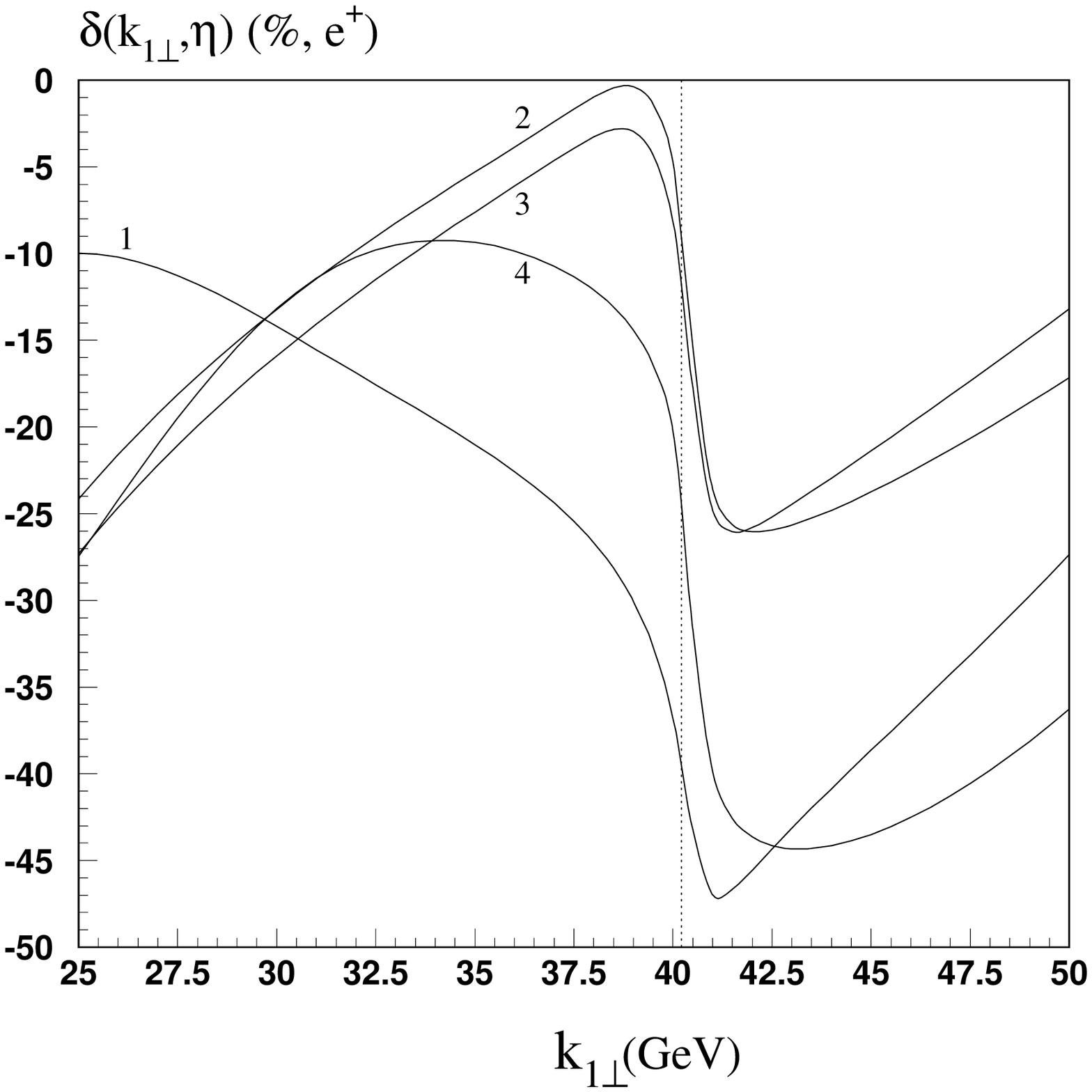} }
\end{picture}
&
\begin{picture}(120,120)
\put(90,0){
\epsfxsize=9cm
\epsfysize=9cm
\epsfbox{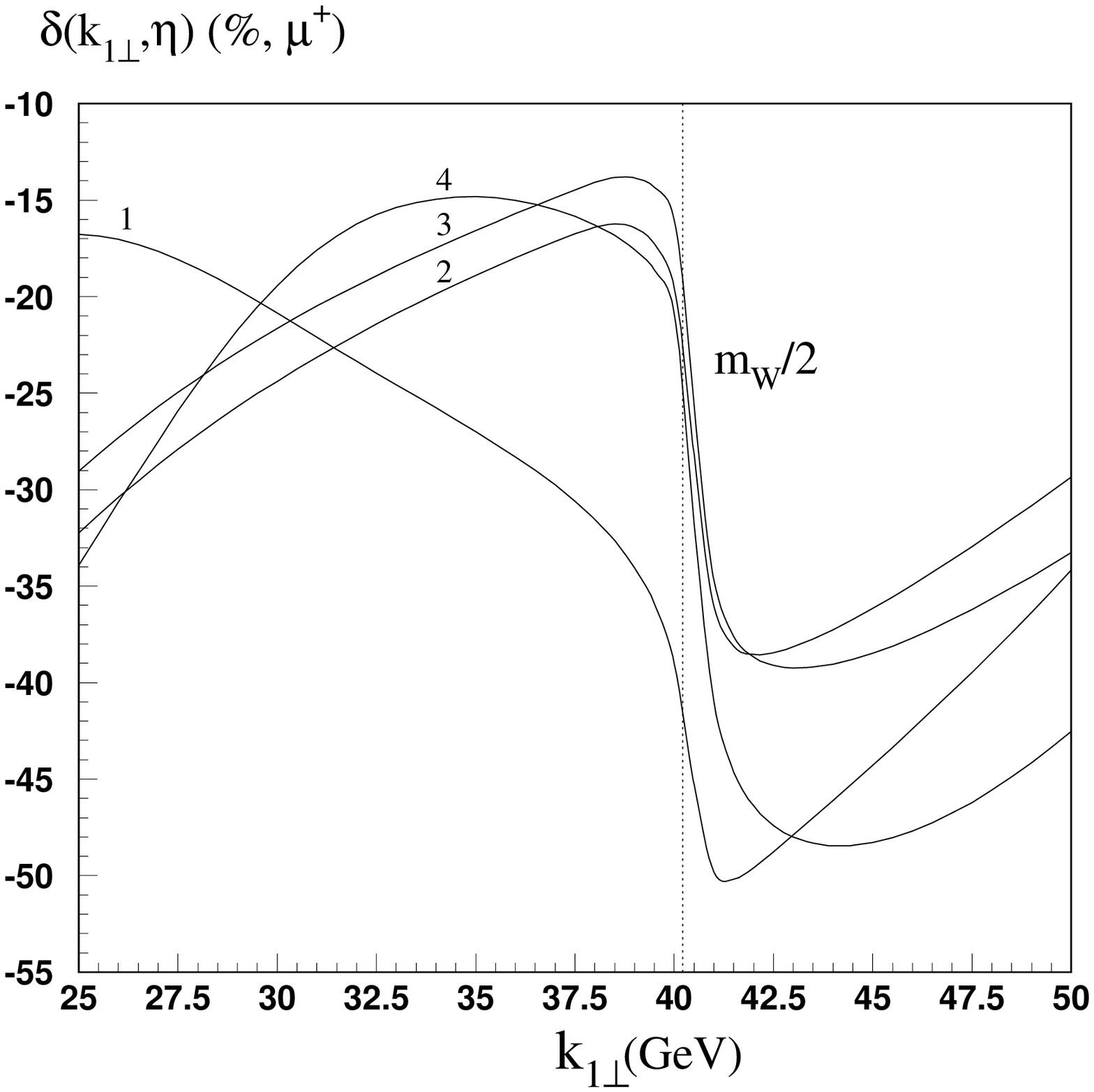} }
\end{picture}
\end{tabular}
\vspace*{-5mm}
\caption{\protect\it
Relative corrections to the lepton transverse momentum distribution
at $\sqrt{S}=14$ TeV, (LHC) for different pseudorapidity $\eta $.
The numbers $1,2,3,4,5,6$ correspond to the lines with
$\eta=-3,-1.3,0,1.3,3,5$
respectively.
The quark masses are defined by (\ref{mm}).
}
\label{lkty}
\end{figure}

To estimate dependence of radiative effects on the leptonic masses
numerically, our plots were constructed for $e^{+ }X$ and $\mu
^{+ }X$ final states.
From the Figs.~\ref{tkty} and \ref{lkty} one can see that
corrections have a positive sign only in the small region of the positron
transverse momentum at Tevatron kinematics
and negative sign for the whole regions of the antimuon transverse
momentum at Tevatron and LHC ones. This situation can be explained
as
suppression of the hard photon emission (whose contribution is
always
positive) by the production of the additional virtual particles
(whose contributions are negative).
The difference in the behavior of EWC for
$e^{+}X$ and $\mu ^{+}X$ final states appears as the contributions
of collinear singularities from the final lepton radiation which
are proportional to $\log (s/m_l^2)$. The other interesting feature
consists in rather high sensibility of the transverse momentum
distribution
to the changes of pseudorapidity. The steep fall of EWC near
$k_{1T}=m_W/2$ comes as the reflection of the resonance behavior.
\begin{figure}[t]
\vspace*{37mm}
\begin{tabular}{cc}
\begin{picture}(120,120)
\put(-20,0){
\epsfxsize=9cm
\epsfysize=9cm
\epsfbox{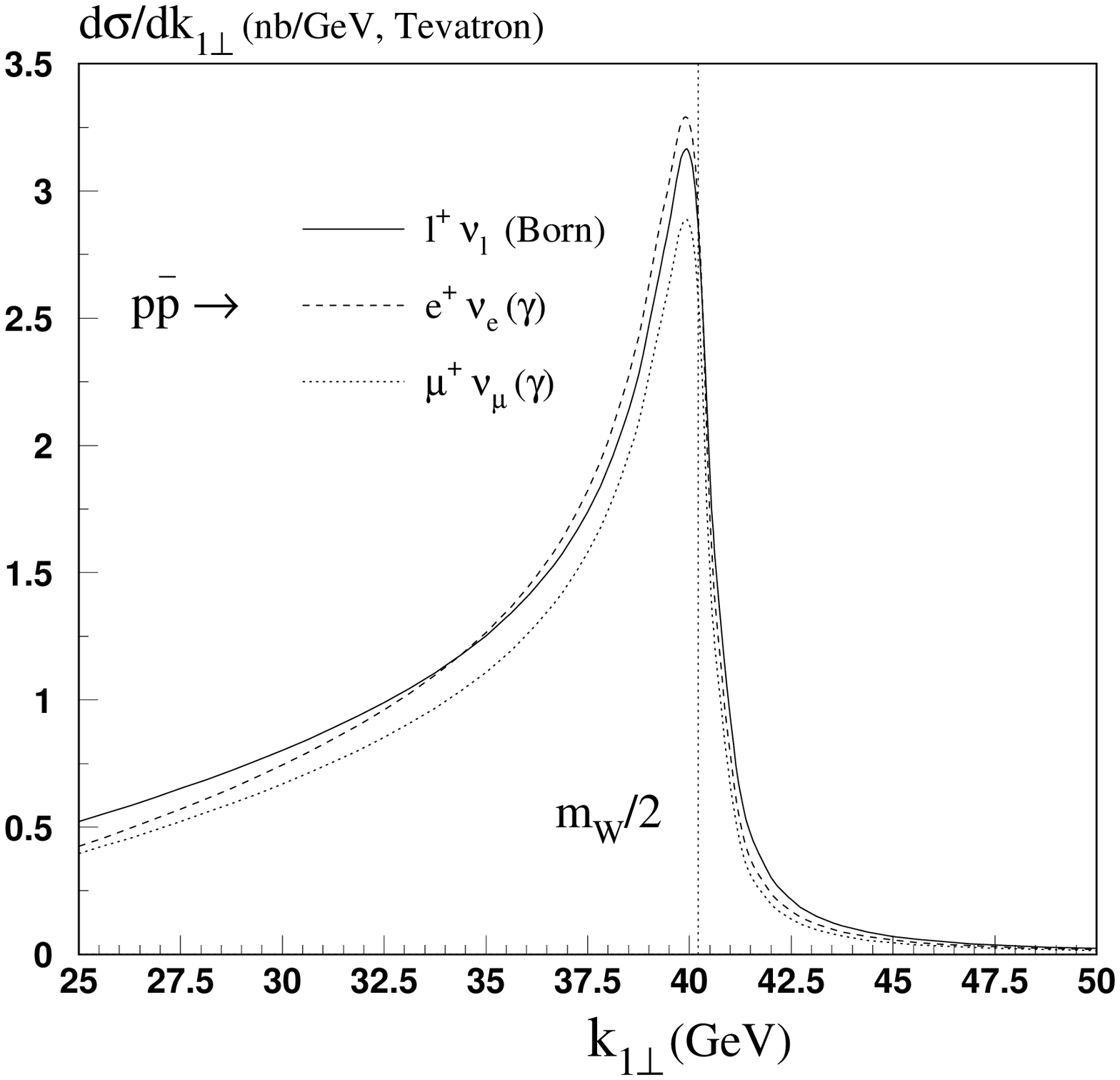} }
\end{picture}
&
\begin{picture}(120,120)
\put(90,0){
\epsfxsize=9cm
\epsfysize=9cm
\epsfbox{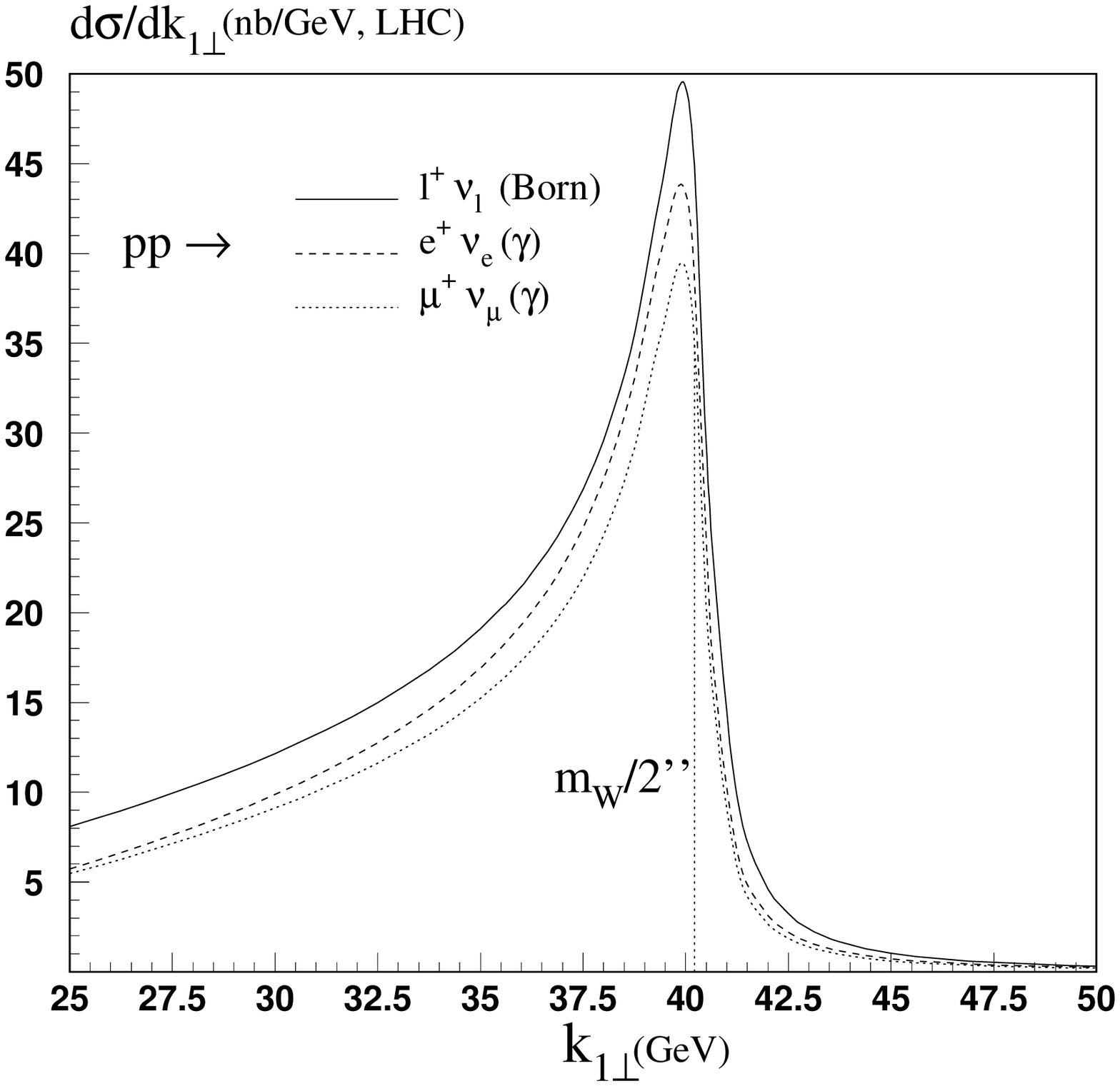} }
\end{picture}
\end{tabular}
\vspace*{-5mm}
\caption{\protect\it
The lepton transverse momentum distribution on the Born level and
with taking into account EWC at
$\sqrt{S}=1.8$ TeV, $|\eta |<1.2$ (Tevatron) and
$\sqrt{S}=14$ TeV, $|\eta |<5$ ( LHC).
The quark masses are defined by (\ref{mm}).
}
\label{3}
\end{figure}
\begin{figure}
\vspace*{25mm}
\begin{tabular}{cc}
\begin{picture}(120,120)
\put(-20,0){
\epsfxsize=9cm
\epsfysize=9cm
\epsfbox{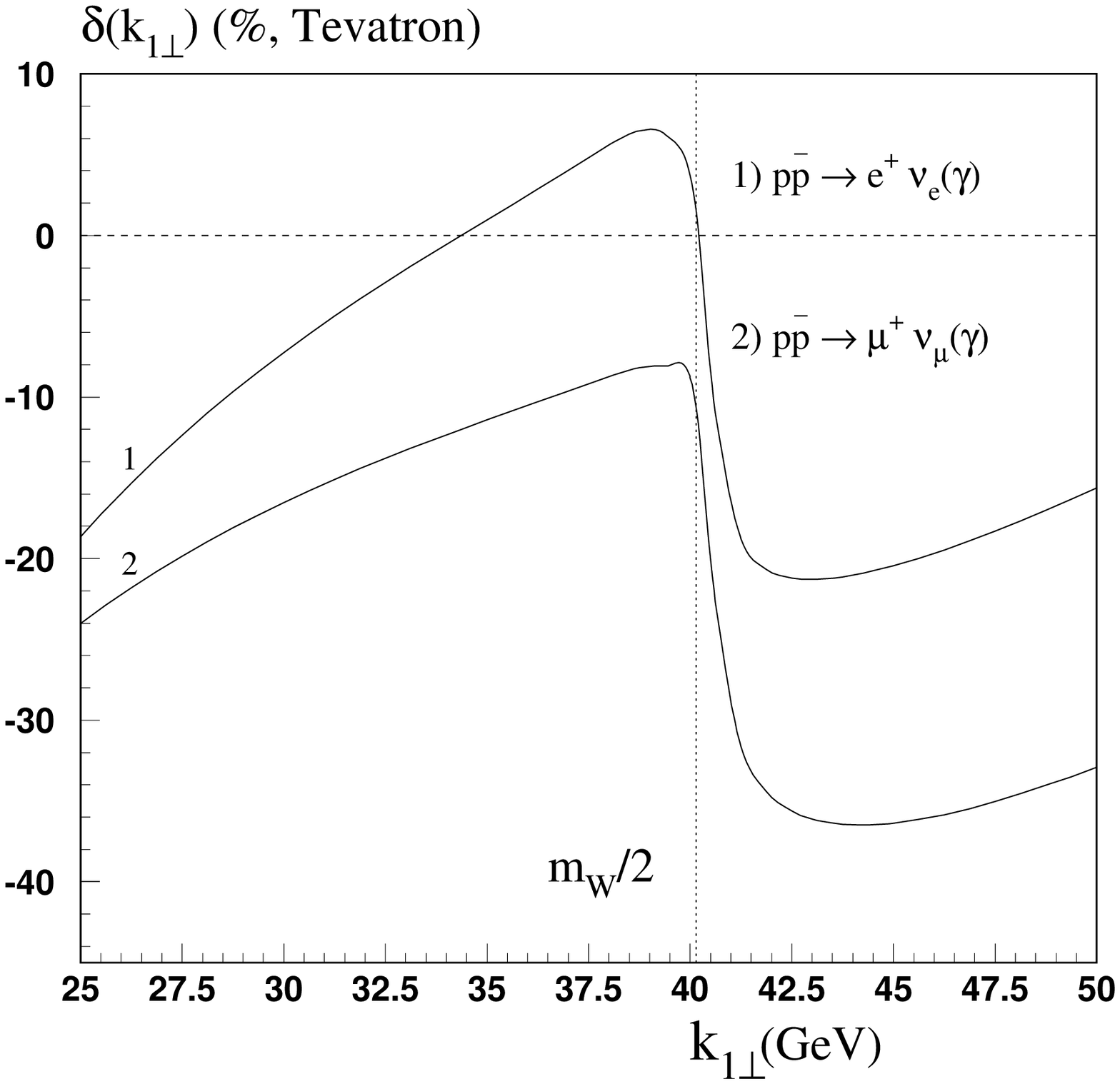} }
\end{picture}
&
\begin{picture}(120,120)
\put(90,0){
\epsfxsize=9cm
\epsfysize=9cm
\epsfbox{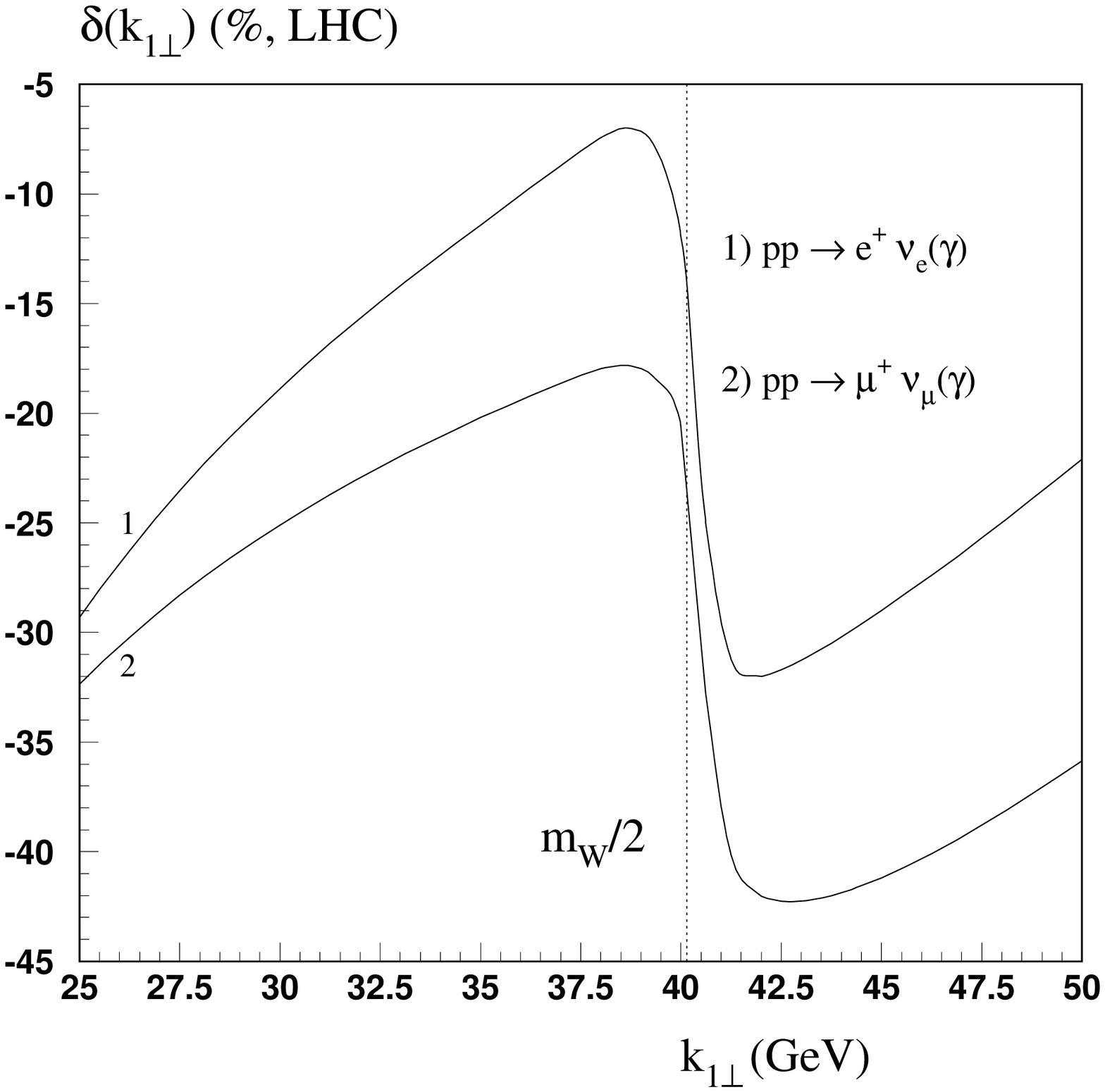} }
\end{picture}
\end{tabular}
\vspace*{-5mm}
\caption{\protect\it
Relative corrections to the lepton transverse momentum distribution
at $\sqrt{S}=1.8$ TeV, $|\eta |<1.2$ (Tevatron) and
$\sqrt{S}=14$ TeV, $|\eta |<5$ (LHC).
The quark masses are defined by (\ref{mm}).
}
\label{5}
\end{figure}

It is shown at Fig. \ref{3} that
after integration over the pseudorapidity ($-1.2<\eta <1.2$ for Tevatron
and  $-5<\eta <5$ for LHC) a positive sign of corrections
\begin{equation}
\delta(k_{1\bot })=\frac
{d\sigma _{RC}/dk_{1\bot}}
{d\sigma _{B}/dk_{1\bot}}
\end{equation}
survives only for the positron final state at Tevatron kinematics
and these corrections reach $\sim 6\%$ at $k_{T1}=39$ GeV.

According to Fig. \ref{5}
EWC decrease the cross section at the peak $k_{1T}=m_W/2$
of the transverse lepton momentum
spectrum by about 11\%  for $\mu ^+$ final state,
increase by about 4\% for $e^+$ final states
for Tevatron,
and decrease the cross section by about 12\% and 24\%
in $e^{+}$ and $\mu^{+}$ final states respectively
for LHC kinematics.

\section{ Conclusions}

Using the covariant approach \cite{covar} and the formulae for
additional virtual particle productions \cite{BS2}
the explicit expressions for the lowest order EWC to the single
$W$-production in hadron-hadron collisions
are obtained.

The main features of our calculation are:
\begin{itemize}
\item
the final expressions for the lowest order EWC
do not depend on any unphysical cutoff
parameters;
\item
during the calculation the quark mass singularity is kept without
any changes;
\end{itemize}

The numerical analysis performed for Tevatron and LHC kinematics
has shown that
\begin{itemize}
\item
EWC are very sensitive to the pseudorapidity changes;
\item
the hard photon emission is essential suppressed
by the additional virtual particle
production. As a result in general EWC to the single $W$-production
has a negative sign almost in the whole region of $W$-experiments
at Tevatron and LHC machines.
\end{itemize}

\end{document}